\documentclass[preprint,3p,authoryear,twocolumn]{elsarticle}
\usepackage{graphicx}
\usepackage{amssymb}
\usepackage{lineno}
\begin{document}
\begin{frontmatter}
\title{Recover of acoustic-gravity wave properties revealed from measurements of VLF radio wave amplitudes }
\author{A. K. Fedorenko}
\author{E. I. Kryuchkov}
\author{O. K. Cheremnykh}
\author{A. D. Voitsekhovska}
\author{Yu. O. Klymenko}
\ead{yurklym@gmail.com}
\cortext[cor1]{Corresponding author.}

\address{$^{\rm a}$Space Research Institute National Academy of Sciences of Ukraine and State Space Agency of Ukraine, prosp. Akad. Glushkova 40, build. 4/1, 03187 MSP Kyiv-187, Ukraine}

\begin{abstract}
A simple method has been proposed to determine acoustic-gravity wave (AGW) properties at mesosphere heights from measurements of VLF radio signals. For relatively short paths (less than $\sim$ 2000 km long), simple relations are obtained for estimation of AGW neutral density fluctuation from measurements of amplitudes of radio signals. In the wave-hop approximation, we relate the fluctuations in radio signal reflection height with AGW-induced vertical displacement of an elementary volume of atmospheric gas. Using these relations, we can recover some properties of AGWs at mesosphere heights.
\end{abstract}
\end{frontmatter}

\section{Introduction}

Acoustic-gravity waves (AGWs) efficiently transfer the energy and the momentum between different atmospheric heights (Hines, 1960; Roy et al., 2019; Cheremnykh et al., 2019). AGWs can also be sources of atmospheric convection and turbulence redistributing the energy of disturbances between different scales (from several thousand kilometers to hundreds of meters). These waves are registered by various satellite and ground-based techniques in the form of periodic ionosphere parameters fluctuations with periods of $\sim$ 5 min up to several hours. According to direct satellite measurements, these disturbances are systematically observed at altitudes of F2 ionosphere region with relative amplitude of several percent (Johnson et al., 1995; Innis, Conde, 2002; Fedorenko, Kryuchkov, 2011; Fedorenko et al., 2015). To study the lower ionosphere, mainly remote methods are used, including a network of ground-based VLF radio stations. Using this method to study the AGWs it is important to relate the observed fluctuations in the amplitudes of VLF radio signals with actual characteristics of the wave disturbances in a neutral atmosphere.

The reflection of VLF radio waves from the upper wall of the Earth-ionosphere waveguide occurs in the daytime at altitudes of $\sim$ 70--74 km (ionosphere D-region) and in the nighttime at altitudes of $\sim$ 85--90 km (lower E-region of ionosphere) (Wait, Spies, 1964; Bezrodny et al., 1988). The interaction of VLF radio waves with the ionosphere is determined primarily by the concentration of electrons, its altitude gradient, and the frequency of electron collisions with neutral particles at the reflection heights. The waveguide geometry and the reflection coefficients are also changed in the presence of the disturbances. It leads to fluctuations of amplitudes and phases in receiving VLF radio signals (Bezrodny et al., 1988). Thus, there is a hypothetical possibility to restore the properties of atmospheric disturbances at the heights of VLF radio wave reflection by measuring their amplitudes. The existing worldwide network of VLF radio stations makes it possible to systematically study the distribution of AGW activity in the lower ionosphere and the mesosphere on a global scale.

Two main approaches are commonly used for a theoretical study of VLF radio wave propagation in the Earth-ionosphere waveguide: the mode theory and the wave-hop theory (Berry, 1964; Yoshida et al., 2008). In the mode theory, the waveguide properties are studied over entire path from the transmitter to the receiver. Here the special software has been developed allowing us to calculate the value and the gradient of the electron concentration depending on the path geometry as well as geo- and heliophysical conditions (Ferguson, 1998). This method is effectively used to study the ionosphere based on the response of VLF radio signals on solar flares (Kolarski, Grubor, 2015). In the wave-hop theory, the amplitude of the radio signal at the receiving point can be considered as a superposition of ground wave with several ionospheric ones which are reflected from the ionosphere (Berry, 1964; Ferguson, 1998; Yoshida et al., 2008).

The AGWs are the local perturbations of the atmosphere with a typical scale -- mainly from tens to hundreds of kilometers. These waves are observed by various methods at different altitudes of the atmosphere and the ionosphere. The amplitudes of AGW-induced fluctuations of VLF radio waves depend on the AGW spectral characteristics, the radio path length, the ionosphere condition direction of the wave propagation, etc. Therefore, it is important to localize the disturbance relative to the path. In this regard, for the study of AGWs, it is advisable to choose relatively short paths (up to $\sim$ 2000 km) for which the wave-hop approximation is valid.

In this paper, we show how some properties of AGWs at mesosphere heights can be recover by using VLF radio signal measurements. The relations connecting the fluctuation of the radio wave amplitudes with AGW neutral density fluctuations are obtained. Using these relations, it is possible to determine the properties of AGW at the heights of the mesosphere.

\section{AGW fluctuations of atmospheric parameters}

AGWs are observed as periodic changes of atmospheric parameters. It is mostly considered in AGW theory the relative fluctuations of atmospheric gas concentration $\Delta N_{n} /N_{n} $, of temperature $\Delta T_{n} /T_{n} $, and of  pressure $\Delta p/p$. These AGW fluctuations are connected by so-called polarization relations (Hines, 1960).

The density AGW fluctuations are related to the displacement of the elementary gas volume by the following relation (Makhlouf et al., 1990):
\begin{equation} \label{1}
\frac{\Delta \rho }{\bar{\rho }} =\left(\frac{\gamma -1}{\gamma } \right)\frac{\Delta z_{n} }{H} +\frac{V_{x} U_{x} }{c_{s}^{2} }.
\end{equation}
Here $\bar{\rho }$ is the undisturbed density, $\Delta z_{n} $ is the vertical displacement of a volume element, $H=kT_n/mg$ is the atmosphere scale height, $\gamma $ is the ratio of specific heats, $U_{x} =\omega /k_{x} $ is the AGW   horizontal phase velocity, $\omega $ is the frequency, $k_{x}$ is the horizontal component of wave vector, $V_{x} $ is the horizontal particle velocity, and $c_{s} $ is the sound speed.

All atmospheric gases are intensively mixed below the turbopause ($\sim$100 km) and, therefore, one can get $\Delta \rho /\bar{\rho }=\Delta N_{n} /\bar{N}_{n} $, where $\bar{N}_{n} $ is the undisturbed neutral particle concentration. For simplification, we consider AGWs with periods $T>T_{B} $ ($T_{B} $ is Brunt-V\"{a}is\"{a}l\"{a} period) and relatively small horizontal phase velocities $V_{x}<<c_{s} $. In this case one can neglect the second term in the right side of Eq. (\ref{1}). We have to also take into account that the temperature in the upper mesosphere decreases with average vertical gradient of $\sim$3${}^\circ$ K/km. The latter can be accounted for by using the additional term which is similar to the non-isothermal frequency of Brunt-V\"{a}is\"{a}l\"{a} (Fedorenko, Kryuchkov, 2014). Under such assumptions, Eq. (\ref{1}) is rewritten as:
\begin{equation} \label{2}
\frac{\Delta N_{n} }{N_{n} } \approx \left(\frac{\gamma -1}{\gamma } +\frac{dH}{dz} \right)\frac{\Delta z_{n} }{H}.
\end{equation}
Here $dH/dz=\left(H/T_{n} \right)\left(dT_{n} /dz\right)$.

Eq. (\ref{2}) relates the fluctuation of the atmospheric gas concentration with the vertical displacement of the volume during an AGW propagation. The physical sense of the equation can be understood from the following considerations. When the wave propagates, the area of adiabatic gas compression moves down until the pressure inside the volume becomes equal approximately the pressure of background atmosphere. However, the gas density inside the volume remains lower than the background density. At first view, this leads to a paradoxical result. Namely, the areas of the adiabatic compression are observed at each altitude level in the form of concentration minima $\left(\Delta N_{n} /N_{n} \right)_{\min } $. But the region of adiabatic gas expansion moves upwards and the density inside the volume remains higher than the background density. Such regions are observed as concentration maxima $\left(\Delta N_{n} /N_{n} \right)_{\max }$ (Dudis, Reber, 1976; Fedorenko, Kryuchkov, 2014).

Due to the high density of the neutral gas in the ionosphere D-region, different chemical reactions involving charged particles occur so quickly that the electron transport can be neglected at the time scales of AGWs  (tens of minutes). One can assume that chemical equilibrium is achieved in the regions of AGW density fluctuations. The velocities of chemical reactions determining the production and the loss of electrons are also changed periodically. Therefore, the AGW neutral density fluctuation will apparent in the form of periodic modulation of electron concentration with the wave period. As a result, AGW propagation at heights of VLF radio wave reflection is accompanied by the periodic fluctuations of the amplitudes and the phases of the signals received (Nina, Cade\u{z}, 2013; Rozhnoi et al., 2014).

\section{Relation of AGW fluctuations with reflection height  of VLF signals }

In the wave-hop approximation, it has been found that AGW-induced fluctuations of the VLF radio wave amplitudes are related to the change of the reflection height by approximate expression (Fedorenko et al., 2019):
\begin{equation} \label{3}
\frac{\Delta A}{\bar{A}} \approx K\Delta h.
\end{equation}
Here $\bar{A}$ is the average undisturbed value of radio wave amplitude, $\Delta A$ is the amplitude disturbance, and $K$ is some function that depends on the path length, the signal frequency, the time of the day, and the conditions in the ionosphere. This function has to be calculated for each radio path separately.

The AGW-induced changes in $\Delta h$ are close in magnitude on different paths and the daily conditions. Periodic fluctuations of the reflection height belong on average to interval of $\Delta h\approx 0.2-0.4km$ (Fedorenko et al., 2019). When AGW propagates in the atmosphere, the vertical displacement $\Delta z_{n} $ of the neutral gas volume can also be from hundreds of meters to several kilometers. Therefore, it is important to relate the $\Delta h$  fluctuations in the reflection height of the VLF radio signals with the AGW properties in the neutral atmosphere.

Let us make some simplifying assumptions. We assume that the volume of the atmospheric gas moves so slowly under AGW propagation that there is enough time for photochemical equilibrium to establish in the volume. The concentration of molecular ions and ion-clusters with which electrons recombine depend primarily on the neutral particle concentration. Therefore, we can assume approximately that the equilibrium density of electrons is determined by the AGW density fluctuation of the neutral particles. We also neglect the influence of AGW temperature fluctuations on value of $\Delta h$.

Further, we suppose that the VLF wave in the absence of disturbances is reflected at some effective altitude level $h$ corresponding to the equilibrium concentration of neutral particles $N_{n0} \left(h\right)$. Hereinafter, the equilibrium concentration will be indicated by subscript "0". A periodic modulation of the neutral density occurs at level $h$ due to the propagation of AGWs. The concentration of the neutral particles equals $N_{n\max } \left(h\right)=N_{n0} \left(h\right)+\Delta N_{n} \left(h\right)$ in the regions of maximums. Here $\Delta N_{n} \left(h\right)$ is the AGW-induced variation of concentration. It is related to the vertical displacement of the volume described by Eq. (\ref{2}) in (Makhlouf et al., 1990).  These regions correspond to upward movement of atmospheric gas volume (see Section 2). The regions of density minima $N_{n\min } \left(h\right)=N_{n0} \left(h\right)-\Delta N_{n} \left(h\right)$ correspond to the volume which is moving down.

The recombination losses of electrons increase in the regions of the neutral concentration maxima. Therefore, the electron concentration at level $h$ becomes insufficient to provide a reflection. The VLF wave may not be reflected at level $h$ but it can be happened at some level $h_{1} =h+\Delta h$ where the same concentrations of the neutral particles and the electrons are achieved as at the unperturbed level. The recombination losses of electrons are less in the regions of concentration minima. In this case, the VLF wave should be reflected at level $h_{2} =h-\Delta h$ where the concentration of the neutral particles is also established as for the unperturbed level.

In view of the foregoing, the condition for the reflection of VLF radio waves in the neutral concentration maxima can be written as
\begin{equation} \label{4}
N_{n0} \left(h_{1} \right)+\Delta N_{n} \left(h_{1} \right)=N_{n0} \left(h\right).
\end{equation}
The equilibrium concentration at level $h_{1} $ is:
\begin{equation} \label{5}
\begin{array}{cc}
N_{n0} \left(h_{1} \right)=N_{n0} \left(h\right)e^{-\frac{\Delta h}{H} \left(1+\frac{dH}{dz} \right)} \approx\\
N_{n0} \left(h\right)\left[1-\frac{\Delta h}{H} \left(1+\frac{dH}{dz} \right)\right].
\end{array}
\end{equation}
From Eqs. (\ref{4}) and (\ref{5}) we can obtain the following expression for relative fluctuations of neutral particle concentration:
\begin{equation} \label{6}
\frac{\Delta N_{n} \left(h_{1} \right)}{N_{n0} \left(h_{1} \right)} =\frac{\Delta h}{H} \left(1+\frac{dH}{dz} \right).
\end{equation}

The condition for the reflection of VLF radio waves in the neutral concentration minima has the form
\begin{equation} \label{7}
N_{n0} \left(h_{2} \right)-\Delta N_{n} \left(h_{2} \right)=N_{n0} \left(h\right).
\end{equation}
Given expression
$$\begin{array}{cc} N_{n0} \left(h\right)=N_{n0} \left(h_{2} \right)e^{-\frac{\Delta h}{H} \left(1+\frac{dH}{dz} \right)} \approx \\ N_{n0} \left(h_{2} \right)\left[1-\frac{\Delta h}{H} \left(1+\frac{dH}{dz} \right)\right]\end{array}$$
for relative concentration fluctuations in the regions of the minima we obtain
\begin{equation} \label{8}
\frac{\Delta N_{n} \left(h_{2} \right)}{N_{n0} \left(h_{2} \right)} =\frac{\Delta h}{H} \left(1+\frac{dH}{dz} \right).
\end{equation}
Equations (\ref{6}) and (\ref{8}) are identical. They relate the AGW amplitude at the reflection height radio waves with shift of $\Delta h$.

From Eqs. (\ref{3}), (\ref{6}), and (\ref{8}) it follows some useful relation between relative fluctuation of neutral density at reflection height and the fluctuation in amplitudes of radio waves $\Delta A/\bar{A}$:
\begin{equation} \label{9}
\frac{\Delta N_{n} }{N_{n0} } =\alpha \frac{\Delta A}{\bar{A}}.
\end{equation}
Here, "transfer" coefficient $\alpha =\left(1+dH/dz\right)/(HK)$ determines by atmosphere parameters $H$ and $dH/dz$ and by function $K$ mostly depending on the path length and the VLF wave frequency (Fedorenko et al., 2019).

From Eq. (\ref{2}) one can obtain the expression for vertical displacement of the neutral gas volume:
\begin{equation} \label{10}
\Delta z_{n} =\frac{\gamma \left(1+dH/dz\right)}{\gamma -1+\gamma dH/dz} \Delta h.
\end{equation}
We have $dH/dz\approx -0.08$ for typical day reflection heights of VLF radio waves (below the mesopause). Reflection of radio waves at night occurs near the mesopause ($dH/dz\approx 0$) or slightly higher ($dH/dz\approx 0.1$). We get $\left(\Delta N_{n} /N_{n} \right)_{day} \approx 3\% $ at daily values of $\Delta h=200m$ and $\left(\Delta N_{n} /N_{n} \right)_{night} \approx 5\% $ for $\Delta h=300m$ at night. Taking into account the temperature dependence on the altitude we get that $\left(\Delta z_{n} /\Delta h\right)_{day} \approx 4.5$ and $\left(\Delta z_{n} /\Delta h\right)_{night} \approx 2.9-3.5$ depending on the gradient value of $dH/dz$ at the reflection height. Therefore, we obtain approximately that $\Delta z_{n} $= 0.6--1.2 km at typical AGW-induced shifts of radio wave reflection level $\Delta h=200-400m$.

For comparison with our results, let us estimate the values of $\Delta z_{n} $ based on the data of AGW observations at the mesosphere heights. The work (Mitchell, Howells, 1998) presents the results of AGW observations at altitudes near the mesopause according to EISCAT radar measurements. Following these investigations, AGWs dominate near the mesopause altitudes with periods of 30--40 min and the amplitudes of vertical velocity of 2.5 m/s. To estimate the vertical displacement of the gas volume we use relation $V_{z} =i\omega \Delta z_{n} $ and obtain $\Delta z_{n} =710-950m$ for AGWs with properties observed. It is in good agreement with our estimates given above. Such AGWs would create fluctuations of reflection level $\Delta h=200-270m$ when observed its using the VLF radio waves.

\section{Conclusions}

It is proposed a simple method which allows us to calculate approximately the fluctuations of the neutral atmosphere caused by AGW propagation from the measurements of the radio signal amplitudes over relatively short paths (less than $\sim$ 2000 km long).

The relations between the AGW amplitudes and the amplitudes of the VLF radio waves are obtained in the wave-hop approximation. Approximate relations (\ref{9}) and (\ref{10}) can be used to determine the fluctuations of the AGW neutral density and the vertical displacement of the gas volume from the measurements of VLF radio waves amplitudes. Thus, we can determine the basic properties of AGW at the heights of the mesosphere.

Note that any information about the properties of the ionospheric plasma has not been used. We directly relate the variations in the effective reflection height with the vertical displacements of the gas volume and the neutral density fluctuations.

The study was supported by grant 2020.02/0015 "Theoretical and experimental studies of global perturbations of natural and artificial origin in the Earth-atmosphere-system" of National Research Foundation of Ukraine.

\end{document}